\documentstyle[twoside,fleqn]{an-art}
\begin{document}
\def\ltsima{$\; \buildrel < \over \sim \;$}
\def\simlt{\lower.5ex\hbox{\ltsima}}
\def\gtsima{$\; \buildrel > \over \sim \;$}
\def\simgt{\lower.5ex\hbox{\gtsima}}
\def\address #1END {{\vspace{9mm}\noindent\small Address of the author: \medskip \\ #1}}
\def\addresses #1END {{\vspace{9mm}\noindent\small Addresses of the authors: \medskip \\ #1}}
\begin{titlepage}
\setcounter{page}{1}
\def\makeheadline{\vbox to 0pt{\vskip -30pt\hbox to 50mm
{\small Astron. Nachr. 1, 1--x \hfill}}}
\makeheadline
\title {First Beppo-SAX results on AGN}
\author{{\sc M. Guainazzi}, Roma, Italy \\
\medskip
{\small A.S.I., Beppo-SAX Science Data Center} \\
\bigskip
{\sc L. Piro}, Frascati, Italy \\
\medskip
{\small Istituto di Astrofisica Spaziale, C.N.R.} \\
}
%
\date{Received 1997 May 20; accepted 1997 July 2} 
\maketitle
%
\summary
In the following paper, some first {\it Beppo-SAX} results on AGN are
presented. Main on-flight calibration features and observational
properties are discussed at the light of possible future AGN
studiesEND
\keyw
AGN - SAx satellite - x-ray observationsEND
\AAAcla
158END
\end{titlepage}
%
%
\kap{The Beppo-SAX scientific payload}

Beppo-SAX (Boella et al., 1997a)
is a joint Italian-Dutch mission aimed at the study of
X-ray emission from celestial objects in the energy range 0.1--300~keV,
with high spatial and good energy and time resolution. The scientific
payload consists of 4 coaligned Narrow Field Instruments (NFI) and a
couple of Wide Field Cameras (WFC), pointing ortogonally to the NFI
direction. Two NFI are proportional counters with
imaging capabilities: the Low Energy Concentrator Spectrometer (LECS), Parmar
et al. 1997, and the Medium Energy Concentrator Spectrometer (MECS), Boella
et al. 1997b. The others are collimated instruments mounted
on a rocking system to achieve a continuous
monitoring of the background along the orbit:
the High Pressure Gas Scintillator Proportional Counter (HPGSPC), Manzo et al.
1997, and the Phoswich Detector System (PDS), Frontera et al. 1997.
In Table~\ref{tab1} the main properties of the NFI are briefly summarized.
\begin{table}
\begin{footnotesize}
\begin{center}
\begin{tabular}{lcccc} \hline
& LECS & MECS & HPGSPC & PDS \\ \hline \hline
Energy band (keV)$^a$ & 0.1--10 & 1.8--10.5 & 7-60 & 20-200 \\
Energy resolution$^b$ & $8.84 \times (E/6$ keV$)^{-0.5}$ &
$8.02 \times (E/6$ keV$)^{-0.5}$ & $11.0 \times (E/6$ keV$)^{-0.5}$ & $15 \times (E/60$ keV$)^{-0.5}$ \\
Encircled energy fraction$^c$& $2.1 \times (E/6$ keV$)^{-0.5}$& ... & ... & ... \\
Effective area ($cm^2$) & 150 @ 6 keV & 50 @ 6 keV & 260 @ 20 keV & 650 @ 60 keV \\ \hline \hline
\end{tabular}
\end{center}
$^a$the energy ranges are quoted for which the response matrices are currently calibrated \\
$^b$$\Delta E/E (\%)$ Full Width Half Maximum (FWHM) \\
$^c$arcmin FWHM
\end{footnotesize}
\caption{Properties of the scientific payload onboard Beppo-SAX}
\label{tab1}
\end{table}
A package for data reduction and analysis and a complete set of
calibrations are available from December 31st 1996. The residuals
on the deconvolved spectrum of the Crab nebula are less than 5\%
for all NFI in the energy ranges reported in Table~\ref{tab1}

In the following paper, the first results obtained by Beppo-SAX on
Active Galactic Nuclei (AGN)
are presented. We will follow a scientifically-driven approach, focusing
some astrophysical issues on which Beppo-SAX is starting to give
interesting insights; the presence of warm absorbing matter in both
Seyfert 1 and blazar objects ($\S 2.1$), the detection of hard ($ E >
20 \ keV$) tails in both radio--quiet and radio--loud AGN ($\S 2.2$).
The Beppo-SAX capability of detecting sub-milliCrab sources
in the mid X-ray band is briefly discussed in $\S 2.3$.

Readers can find more information on Beppo-SAX and its first scientific
results browsing from the URL page {\verb!http://www.sdc.asi.it!}

\kap{First scientific results on AGN}

Beppo-SAX Science Verification Phase (SVP) took place from July to
September 1996 (but a couple of targets which were observed two
months later). More than 30 targets were observed;
among them 4 bright AGN:
the Seyfert MCG-6-30-15 and NGC4151, the core-jet
quasar 3C273 and the blazar PKS2155-304. In Table~\ref{tab2} the log
\begin{table}
\begin{footnotesize}
\begin{center}
\begin{tabular}{lcccccl} \hline
Object & Start Time (UT) & End Time (UT) & $T_{exp}$ (s) & MECS count rate (s$^{-1}$) & PDS count rate (s$^{-1}$) & Reference$^b$ \\ \hline \hline
3C273 & 18-Jul-96 02:03:25 & 21-07-96 08:35:00 & 132000 & $1.084\pm0.003$& $1.09\pm0.03$ & 1 \\
MGC-6-30-15 & 29-Jul-96 18:49:46 & 03-Aug-96 03:15:00 & 185000 & $0.869\pm0.002$ & $0.39\pm0.02$ & 2,3 \\
NGC4151$^a$& 06-Jul-96 16:37:34& 10-Jul-96 04:12:39 & 71840 & $1.516\pm0.004$ & $2.08\pm0.03$ & 4 \\
PKS2155-304 & 20-Nov-96 00:15:58 & 22-Nov-96 13:30:06 & 107700 & $1.089\pm0.003$ & $0.20\pm0.03$ & 5 \\ \hline \hline
\end{tabular}
\end{center}
$^a$ a second observation has been performed five months later for a total
exposure time $\sim 35000 \ s$ to complete the originally scheduled 100 ks \\
$^b$1. Grandi et al., 1997, A\&A, submitted \\
2. Orr et al., 1997, A\&A, submitted \\
3. Molendi et al., 1997, A\&A, submitted \\
4. Piro et al., 1997, A\&A, submitted \\
5. Giommi et al., 1997, A\&A, submitted
\end{footnotesize}
\caption{AGN observed during Beppo-SAX SVP. MECS exposure times are reported.}
\label{tab2}
\end{table}
of such observations is reported. The papers quoted in table contains a
preliminary analysis of the data and the reader should refer to them
for any result regarding the individual sources.

\sect{Warm absorber: an ubiquitous feature in AGN?}

Recent ASCA results have demonstrated narrow absorption features by highly
ionized matter to be a common feature in Seyfert 1 spectra (Reynolds
et al., 1997; Mushotzky et al., 1997). The MCG-6-30-15 Beppo-SAX
low-energy spectrum is consistent with a warm absorber scenario
where highly ionized specied of Oxygen and Neon are responsible for
the most prominent absorption features at $E \simlt 1 \ keV$.
Typical numerical density ion ratio $n_{OVII}/n_{OVIII}$ range
from 0.2 to 1.7, corresponding to a thermal equilibrium photoionzed
plasma by an AGN-like continuum if the ionization parameter $\xi \sim
40 \div 70 \ erg \ s^{-1} \ cm^{-1}$ and $T \sim$ few $10^5 \ K$.
However, the low energy light curves
display a pattern of variability that cannot easily be explained by
simple one-zone photoionization-driven warm absorber in thermal
equilibrium.
Figure~\ref{fig1} shows the light curves
\begin{figure}
\vspace{60mm}
\caption{Light curves
from the Beppo-SAX observation of MGC-6-30-15
in the 0.7--1.5~keV ({\it upper panel})
and 1.5--2.5~keV energy ranges ({\it middle panel}) and
hardness ratio (HR, {\it lower panel}) (binning time 1000 s).
Dotted and dash-dotted lines in the lower panel are 1 $\sigma$
and 3 $\sigma$ around the mean $\bar{HR}=1.67$}
\label{fig1}
\end{figure}
in the 0.7--1.5~keV and 1.5--2.5~keV energy bands. The former (LECS)
should be the most affected by any ionized absorbed,
while the latter (MECS) is likely to be dominated by the bare
nuclear continuum. Let's consider
the first 100 ks, which are characterized by a
quasi-symmetric flux modulation by a factor $\sim$ 5. 
A sharp rise of the hard flux (two-fold time $\sim 10000 \ s$)
is followed by the soft flux with a delay $\Delta t \sim 20000 \ s$
and the HR value is therefore significantly different from
the mean measured along the whole observation.
The soft peak appears also smoother then the hard, suggesting a
longer stay time in the high state, although the not even
sampling could be partly responsible for such an effect.
Such findings
suggest the ionization and recombination timescales
to be typically longer than the
variability timescales in such a source. That makes difficult to
think that photoionization equilibrium is always effective. We
stress moreover that no significant variation of the HR is present
in the rest of the observation, despite flux variation wider than
5.
 
Similar absorption structures are starting to be detected
in the low energy spectrum of other kinds of AGN,
where they had not been observed yet, or show a much
more complex structure than previuosly seen. In 3C273 a
narrow-band absorption feature
at $E_{observer} \simeq$ 0.5--0.6~keV, has been
unambiguously detected for the first time
(see Figure~\ref{fig2}). It can be phenomenologically
\begin{figure}
\vspace{60mm}
\caption{Energy vs. optical depth ({\it left}) and covering fraction
({\it right}) of the edge and saturated line model
of the absorption structure in the low-energy spectrum of 3C273. The
observer energies of an {\sc O vii} edge and Ly-{$\alpha$}
{\sc O viii} notch at the source reshift are shown for comparison}
\label{fig2}
\end{figure}
modelled with either with
a photoionization absorbption edge or with a saturated absorption
line. In the former scenario, the rest frame energy is consistent with
an average Oxygen ionization stage from {\sc O iii} to {\sc O v}.
An explanation of such a feature as a comoving warm absorbing matter
with the host galaxy can be ruled out since
stronger continuum opacity and absorption features from moderate
ionized elements like {\sc C vi}, {\sc C iv} or helium
should be detected in the spectrum as
well.
An MCG-6-30-15-like warm absorber would require
an inflowing matter with bulk velocity $\sim 60000-90000 \ km \ s^{-1}$.
Alternatively, the most
likely candidate for a saturated absorption is the resonant
Ly-${\alpha}$ absorption in {\sc O viii}. In such a case the observed
energy can be reconciled with the laboratory one if the absorption
occur in a matter with a bulk velocity {\it towards} the observer $\sim
60000 \ km \ s^{-1}$,
and could be therefore associated with the outflowing
jet already seen at radio-optical wavelengthts.
Both such extreme interpretations are however quite puzzling.

PKS2155-304 is a BLLac object with a very steep X-ray
spectrum in the mid X-ray ($\Gamma_{3-10 keV} \simeq 2.6$).
EXOSAT (Madjeski et al. 1991) and BBRXT (Madjeski et al. 1997)
observations have already
showed a notch structure at rest frame energy consistent with a resonant
Ly-${\alpha}$ absorption from {\sc O viii}. The Beppo-SAX spectrum, extending
in the low-energy band down to
0.1~keV, has revealed instead a much more complex spectrum;
the notch feature is likely to be accompanied by a further absorption edge
at $E_{observer} \sim 0.22 \ keV$ or an even more complex multicomponent
edge structure could well model the observational data.

\sect{Hard tails}

The very low in-flight background in the PDS detector has allowed the
detection of hard tail emission in the 20--100~keV from several AGN. In
NGC1068 (Matt et al. 1997) it has been the first
detection reported so far,
and has allowed an independent confirmation
of the
predominance of the cold reflector in the
double-mirror scenario that had been already invoked to explain
the multicomponent profile of the iron line K$_{\alpha}$ fluorescence line
(Iwasawa et al. 1997).
An analogously important case is PKS2155-304, where the hard tail
has been for the first time reobserved after the first putative detection
by Urry et al. (1982). In this case the
emission is likely to be associated with inverse
Compton scattering of the syncrotron radiation and confirm the
interpretative unified scenario for the blazar spectral energy
distribution as suggested by Padovani $\&$ Giommi (1995).

The net count rates in the PDS are $0.20 \pm 0.03$ and $0.12 \pm 0.03$
for PKS2155-304 and NGC1068 respectively, corresponding to a 3.5 $\sigma$
detection in the latter case. In both cases the associated flux
in the 20--100~keV band is $\sim$ 1-2 mCrab.
Such a low flux level corresponds to
a less than $\simeq 1\%$ of the net count rate in the PDS
units. A very strong control of the systematics
associated with the background subtraction procedure is
needed. PDS sensitivity
has been widely investigated during the SVP (Guainazzi \& Matteuzzi 1997).
In Figure~\ref{fig3} a blank sky
\begin{figure}
\vspace{60mm}
\caption{Blank sky PDS spectrum ($T_{exp} \sim 225 \ ks$). Two mCrab
power-law spectra with photon index $\Gamma=1.5$ ({\it dotted
line}) and  $\Gamma=2$ ({\it dashed-dotted line}) are superimposed
for comparision (from Guainazzi $\&$ Matteuzzi 1997)}
\label{fig3}
\end{figure}
PDS spectrum accumulated on $\sim 225 \ ks$, is shown, and 1 mCrab
power-law spectra with spectral photon index $\Gamma=1.5$ and
$\Gamma = 2$ are superimposed
for comparision. The residual systematics are well below the simulated
spectrum up to $E \simeq 50 \ keV$. Such a striking sensitivity can be achieved
due to the good reproducibility of the instrumental background on timescales
as low as 100 seconds. mCrab sources with moderate soft spectra can be
therefore
confidently measured by Beppo-SAX, with a much higher sensitivity than
any analogous past and/or foreseeable in the next future mission.
A refinement of the selection criteria in the nearby future will
probably allow to achieve sensitivities up to few fraction of mCrab.

The broad-band coverage of Beppo-SAX instruments has also allowed the first
robust detection of a Compton reflection component in the intermediate
Seyfert galaxy NGC4151. In Figure~\ref{fig4} the spectrum in the
\begin{figure}
\vspace{60mm}
\caption{3.5--200~keV Spectrum ({\it upper panel}) and residuals in units of
data/model ratio ({\it lower panel}) when a simple power-law model
is applied to the data of the intermediate Seyfert
galaxy NGC4151. Note the ``emission bump'' between 20 and 50~keV}

\label{fig4}
\end{figure}
whole 3.5--200~keV is shown when a simple power-law model with
index fixed to the intermediate ({\it i.e.:} MECS) band is used.
The spectral points between 20 and 50 keV
lay well above the extrapolation of the intrinsec
continuum, in analogy to what already observed in most Seyfert 1s
(Pounds et al. 1990, Matsuoka et al. 1990, Nandra \& Pounds 1994).
The ratio between the reflected and direct component normalizations
ranges betweem 13\% and 40\% and appears not to be strictly
correlated with the highly variable (more than a factor 2) nuclear
flux, either during the first long ($\simeq$ 3 days) observation
or between the two observations performed, which were several
months apart; such an evidence suggest any reprocessing matter to
be distant at least $\simeq 350 \ AU$ from the central engine.

\sect{Sub-millicrab AGN}

The MECS detector has shown a very low and stable background level.
For comparision, the GIS background level is $\sim$ a factor of 2 higher.
Such a feature has a strong impact on the determination
of the spectral properties of sub-milliCrab AGN. Currently at least two
Guest Observer programs are ongoing that take full advantage of such a
feature. Beppo-SAX has been observing a sample of Narrow Line Seyfert 1
Galaxies with exposure times 20--40 ksec. The first results (Comastri et
al. 1997) allow a much better determination of the spectral shape
in the $E > 2 \ keV$ band then in the past, revealing evidence of
spectral curvature around $E \simeq 3 \ keV$, and
$\Gamma_{hard}$ slightly higher or marginally consistent with
typical Seyfert 1 ones. On the other hand,
the first results from a sample of optically selected
Seyfert 2 galaxies according to their {\sc O [iii]} luminosity has
produced a 100\% detection rate (Salvati et al. 1997a and 1997b
in preparation), and several sigma detections of
Compton-thick (thin) objects as faint as $\sim 10^{41} \ erg \ s^{-1}$
($\sim 10^{40} \ erg \ s^{-1}$) can be achieved.

\kap{What Beppo-SAX can provide to future AGN studies}

We can therefore summarize the best features that Beppo-SAX can provide
to the international community of AGN researchers as follow:

\begin{itemize}

\item broadband spectral capability in the whole 0.1--200~keV, with
imaging capability at the highest resolution ever flown in the mid-hard
X-ray band and good spectral resolution

\item high quality calibration (residuals less than 5\% on the deconvolved
Crab spectrum for all NFI)

\item unprecedented low background both in the intermediate and in the
hard band. Targets as faint as a fraction of mCrab (in the 2--10~keV
energy range) or $\sim$ mCrab in the 20--100~keV range can be
detected at several sigma level of significancy and reliable determination
of flux and rough spectral shape is at hand

\item increasing availability of time allocation for guest observers'
proposals (40\% of next Annonuncement of Opportunity time will be open to
the international community)

\end{itemize}

%
\refer
\aba
\rf{Boella G., Butler R.C., Perola G.C. et al., 1997a, A\&AS, 122,}
\rf{Boella G., Chiappetti L., Conti G., et al., 1997b,  A\&AS, 122, 327}
\rf{Comastri A., et al., 1997, A\&A, submitted}
\rf{Frontera F., Costa E., Dal Fiume D., et al., 1997, A\&A, 122, 357}
\rf{Guainazzi M., Matteuzzi A., SDC-TR-001,
available at the URL:{\verb!http://www.sdc.asi.it/software/cookbook/pds.html!}
\rf{Iwasawa K., Fabian A.C., Matt G., 1997, MNRAS, in press}
\rf{Madejsky G.M., Kallmann T., Serlemitos P.J., et al., 1997, ApJ,
submitted}
\rf{Madejsky G.M., Mushotzky R.F., Weaver K.A., et al., 1991, ApJ, 370,
198}
\rf{Manzo G. Giarrusso S., Santangelo A., et al., 1997, A\&A, 122}
\rf{Matsuoka M., Piro L., Yamauchi M., Murakami T., 1990, ApJ, 361, 440}
\rf{Matt G., Guainazzi M., Frontera F., et al., 1997, A\&A, submitted}
\rf{Mushotzky R.F., Proceeding of the Workshop ``Mass Ejection from AGN'',
Pasadena, 1997, in press}
\rf{Nandra K., Pounds K., 1994, MNRAS, 268, 405}
\rf{Padovani P., Giommi P., 1996, ApJ, in press}
\rf{Parmar A., Martin D.D.E., Bavdaz M., et al., 1997, A\&A, 122, 309}
\rf{Pounds K., Nandra K., Stewart G.C., et al., 1990, Nature, 344, 132}
\rf{Reynolds C.S., Proceeding of the Workshop ``Mass Ejection from AGN'',
Pasadena, 1997, in press}
\rf{Salvati M., Bassani L., Maiolino R., et al., 1997a, A\&A, in press}
\rf{Urry C.M., Mushotzky R.F., 1982, ApJ, 253, 38}

\abe

%
\address
Matteo Guainazzi\\
A.S.I., Beppo-SAX Science Data Center\\
Via Corcolle 19\\
I-00131 Roma\\
ItalyEND
%

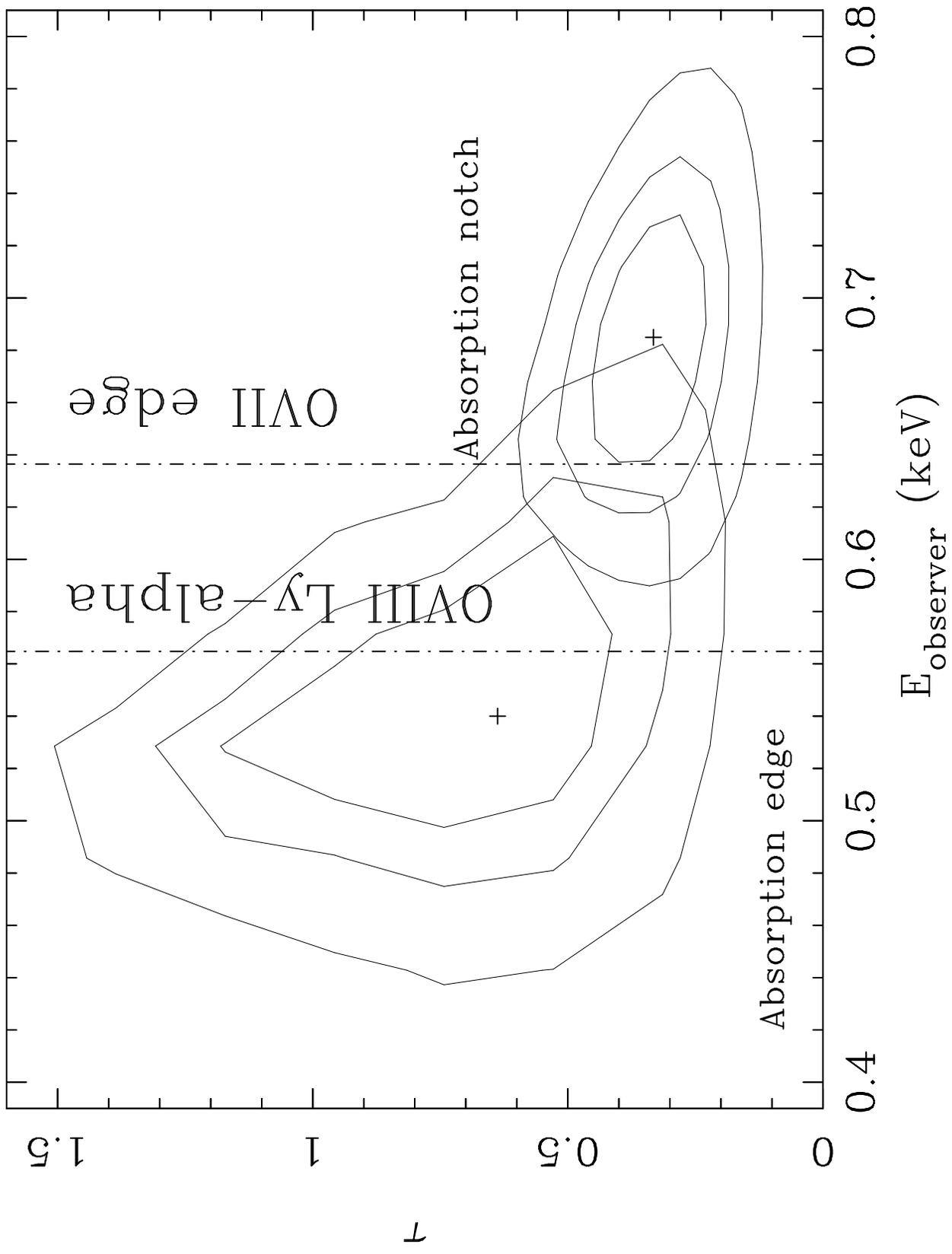
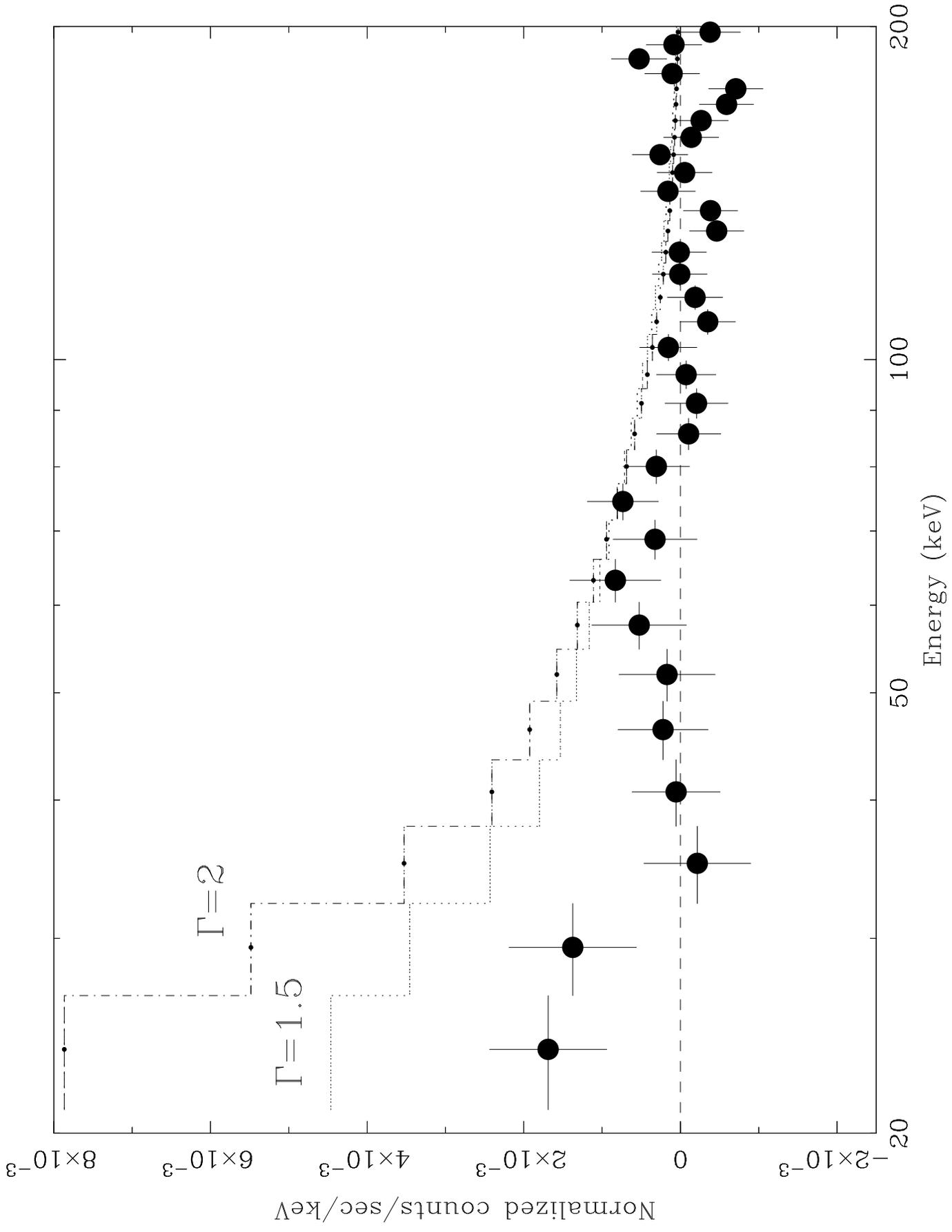
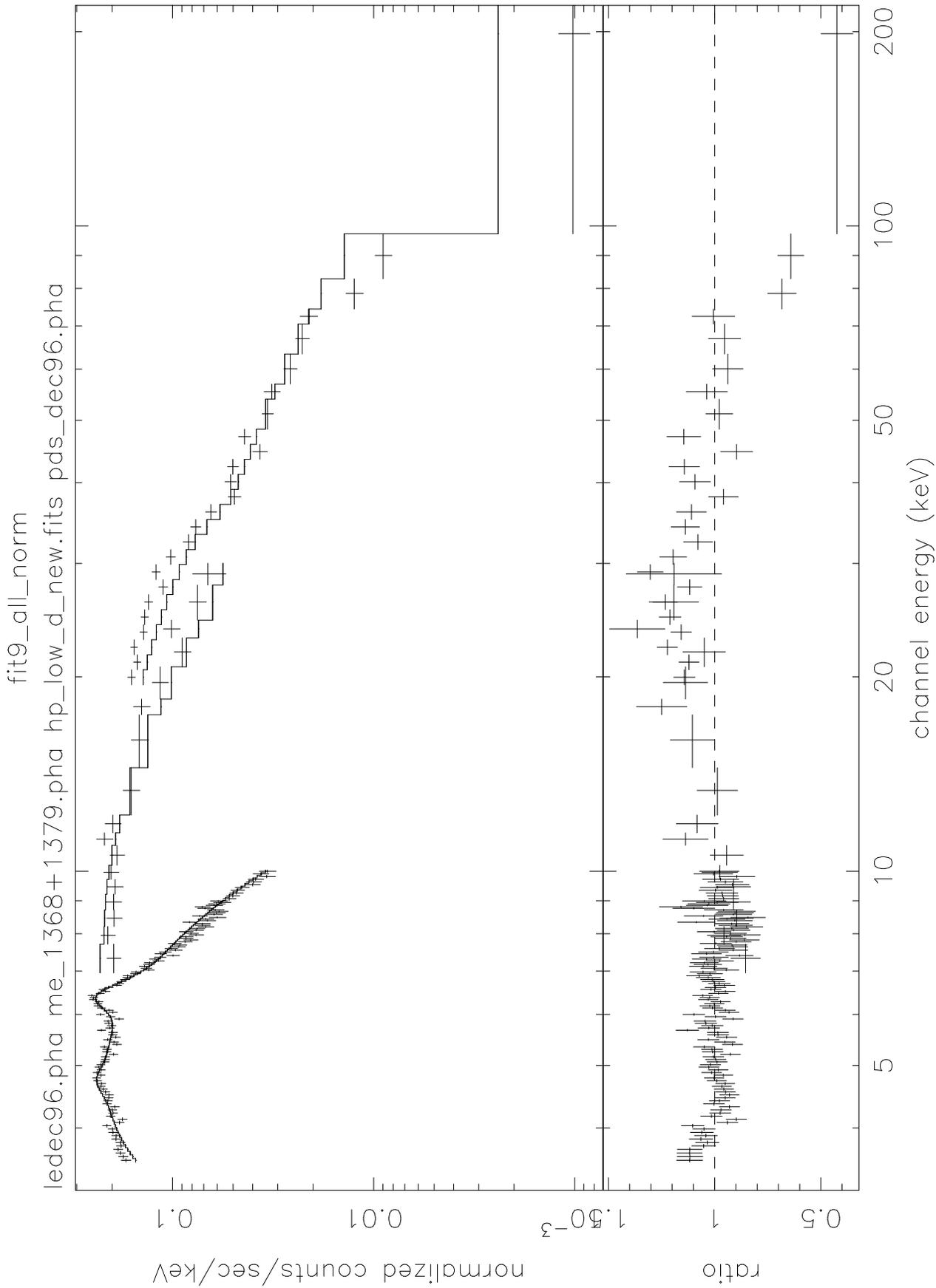
\end{document}